%% file: article.tex
\documentclass{amsart}
\usepackage{amsmath,amssymb,amsthm,graphicx,hyperref,srcltx,changepage,subfig,color}
\usepackage[capitalize,nameinlink]{cleveref} 
\crefname{approach}{Condition}{conds}

\newcommand\bbR{\mathbb{R}}
\newcommand\bbN{\mathbb{N}}
\newcommand\bbG{\mathbb{G}}
\newcommand\bbH{\mathbb{H}}
\newcommand\bxi{\boldsymbol{\xi}}
\newcommand\bx{\boldsymbol{x}}

\newcommand\bQ{\boldsymbol{Q}}

\newcommand\bn{\boldsymbol{n}}

\newcommand\bA{{\bf A}}
\newcommand\bB{{\bf B}}
\newcommand\bD{{\bf D}}
\newcommand\bM{{\bf M}}

\newcommand\dd{\,\mathrm{d}}
\newcommand\He{\mathit{He}}

\newcommand\bw{\boldsymbol{w}}

\newcommand\bv{\boldsymbol{v}}

\newcommand\mH{{\mathcal H}}
\newcommand\mP{{\mathcal P}}
\newcommand\bmH{\boldsymbol{\mH}}

\newcommand\pd[2]{\dfrac{\partial {#1}}{\partial {#2}}}

\newcommand\comment[1]{}

\graphicspath{{images/}}
{\theoremstyle{remark} }

\newtheorem{definition}{Definition}
\newtheorem{approach}{Condition}

\title[Hyperbolic Model Reduction]{Hyperbolic Model Reduction for
  Kinetic Equations}

\author[Z.-N. Cai]{Zhenning Cai} \address[Zhenning Cai]{Department of
  Mathematics, National University of Singapore, 10 Lower Kent Ridge
  Road, Singapore 119076} \email{matcz@nus.edu.sg}

\author[Y.-W. Fan]{Yuwei Fan} \address[Yuwei Fan]{Department of Mathematics, Stanford University, Stanford, CA 94305.}
\email{ywfan@stanford.edu}

\author[R. Li]{Ruo Li} \address[Ruo Li]{CAPT, LMAM and School of
  Mathematical Sciences, Peking University, Beijing 100871,
  P.R. China} \email{rli@math.pku.edu.cn}

\begin{document}

\begin{abstract}
  We make a brief historical review of the moment model reduction for
  the kinetic equations, particularly Grad's moment method for
  Boltzmann equation. We focus on the hyperbolicity of the reduced
  model, which is essential for the existence of its classical
  solution as a Cauchy problem. The theory of the framework we
  developed in the past years is then introduced, which preserves
  the hyperbolic nature of the kinetic equations with high
  universality. Some lastest progress on the comparison between models
  with/without hyperbolicity is presented to validate the
  hyperbolic moment models for rarefied gases.
\end{abstract}

\maketitle

\input{review}
\input{framework}
\input{latest}
\input{conclusion}

\bibliographystyle{plain}
\bibliography{../article}

\end{document}

%% file: review.tex
\section{Historical Overview}

The moment methods are a general class of modeling methodologies for 
kinetic equations. We would like to start this paper with a historical
review of this topic. However, due to the huge amount of references, a
thorough overview would be lengthy and tedious.  Therefore, in this
section, we only restrict ourselves to the methods related to the
hyperbolicity of moment models. Even so, our review in the following
paragraphs does not exhaust the contributions in the history.

According to Sir J.H.Jeans \cite{jeans1967introduction}, the kinetic
picture of a gas is ``a crowd of molecules, each moving on its
own independent path, entirely uncontrolled by forces from the other
molecules, although its path may be abruptly altered as regards both
speed and direction, whenever it collides with another molecule or
strikes the boundary of the containing vessel.'' In order to describe
the evolution of non-equilibrium gases using the phase-space
distribution function, the Boltzmann equation was proposed
\cite{Boltzmann} as a non-linear seven-dimensional partial
differential equation. The independent variables of the distribution
function include the time, the spatial coordinates, and the velocity.

In most cases, the full Boltzmann equation cannot be solved even
numerically. One has to characterize the motion of the gas by
resorting to various approximation methods to describe the evolution
of macroscopic quantities. One successful way to find approximate
solutions is the Chapman-Enskog method \cite{Chapman, Enskog}, which
uses a power series expansion around the Maxwellian to describe
slightly non-equilibrium gases. The method assumes that the
distribution function can be approximated up to any precision only
using equilibrium variables and their derivatives. Alternatively,
Grad's moment method \cite{Grad} was developed in the late 1940s. In
this method, by taking velocity moments of the Boltzmann equation,
transport equations for macroscopic averages are obtained. The
difficulty of this method is that the governing equations for the
components of the $n$-th velocity moment also depend on components of
the $(n+1)$-th moment. Therefore, one has to use a certain closing
relation to get a closed system after the truncation.

Among the models given by Grad's method \cite{Grad}, Grad's 13-moment
system is the most basic one beyond the Navier-Stokes equations, as
any Grad's models with fewer moments do not include either stress
tensor or heat transfer. In \cite{Gombosi1994}, it was commented that
Grad's moment method could be regarded as mathematically equivalent to
the Chapman-Enskog method in certain cases. Thus the deduction of
Grad's 13-moment system can be regarded as an application of
perturbation theory to the Boltzmann equation around the equilibrium.
Therefore, it is natural to hope that the 13-moment system will be
valid in the vicinity of equilibrium, although it was not expected to
be valid far away from the equilibrium distribution \cite{Grad1952}.
However, due to its complex mathematical expression, it is even not
easy to check if the system is hyperbolic, as pointed out in
\cite{Jabin2015KRM}. As late as in 1993, it was eventually verified in
\cite{Muller1993, Muller} that the 1D reduction of Grad's 13-moment
equations is hyperbolic around the equilibrium.

In 1958, Grad wrote an article ``Principles of the kinetic theory of
gases'' in Encyclopedia of Physics \cite{Grad1958}, where
he collected his own method in the class of ``more practical expansion
techniques''. However, successful applications of the 13-moment system
had been hardly seen within two decades after Grad's classical paper
in 1949, as mentioned in the comments by Cercignani
\cite{Cercignani1969}. One possible reason was found by Grad himself
in \cite{Grad1952}, where it was pointed out that there may be
unphysical sub-shocks in a shock profile for Mach number greater than
a critical value. However, the appearance of sub-shocks cannot give
any hints on the underlying reason why Grad's moment method does not
work for slow flows. Nevertheless, Grad's moment method was still
pronounced to ``open a new era in gas kinetic theory''
\cite{Harris1971}.

In our paper \cite{Grad13toR13}, it was found astonishingly that in
the 3D case, the equilibrium is NOT an interior point of the
hyperbolicity region of Grad's 13-moment model. Consequently, even if
the distribution function is arbitrarily close to the local
equilibrium, the local existence of the solution of the 13-moment
system cannot be guaranteed as a Cauchy problem of a first-order
quasi-linear partial differential system without analytical data. The
defects of the 13-moment model due to the lack of hyperbolicity had
never been recognized as so severe a problem. The absence of
hyperbolicity around local equilibrium is a candidate reason to
explain the overall failure of Grad's moment method. 

After being discovered, the lack of hyperbolicity is well accepted as
a deficiency of Grad's moment method, which makes the application of
the moment method severely restricted. ``There has been persistent
efforts to impose hyperbolicity on Grad's moment closure by various
regularizations'' \cite{ZhaoWF2016}, and lots of progress has been
made in the past decades. For example, Levermore investigated the
maximum entropy method and showed in \cite{Levermore} that the moment
system obtained with such a method possesses global hyperbolicity.
Unfortunately, it is difficult to put it into practice due to the lack
of a finite analytical expression, and the equilibrium lies on the
boundary of the realizability domain for any moment system containing
heat flux \cite{Junk}. Based on Levermore's 14-moment closure, an
affordable 14-moment closure is proposed in \cite{McDonald} as an
approximation, which extends the hyperbolicity region to a great
extent. Let us mention that actually in \cite{Grad13toR13}, we also
derived a 13-moment system with hyperbolicity around the equilibrium.

It looks highly non-trivial to gain hyperbolicity even around
the equilibrium, while things changed not long ago. Besides the
achievement of local hyperbolicity around the equilibrium, the study
on the globally hyperbolic moment systems with large numbers of
moments was also very successful in the past years. In the 1D case
with both spatial and velocity variables being scalar, a globally
hyperbolic moment system was derived in \cite{Fan} by regularization.
Motivated by this work, another type of globally hyperbolic moment
systems was then derived in \cite{Koellermeier} using a different
strategy. The model in \cite{Fan} is obtained by modifying only the
last equation and the model in \cite{Koellermeier} revises only the
last two equations in Grad's original system. The characteristic
fields of these models (genuine nonlinearity, linear degeneracy, and
some properties of shocks, contact discontinuities, and rarefaction
waves) can be fully clarified, as shows that the wave structures are
formally a natural extension of Euler equations.

In \cite{Fan_new}, the regularization method in \cite{Fan} is extended
to multi-dimensional cases. Here the word ``multi-dimension'' means
that the dimensions of spatial coordinates and velocity are any
positive integers and can be different. The complicated
multi-dimensional models with global hyperbolicity based on a Hermite
expansion of the distribution function up to any degree were
systematically proposed in \cite{Fan_new}. The wave speeds and the
characteristic fields can be clarified, too. Later on, the
multi-dimensional model for an anisotropic weight function with global
hyperbolicity was derived in \cite{ANRxx}.

Achieving global hyperbolicity was definitely encouraging, while it
sounded like a huge mystery for us how the regularization worked in
the aforementioned cases. Particularly, the method cannot be applied
to moment systems based on a spherical harmonic expansion
of distribution function such as Grad's 13-moment system. As we
pointed out, the hyperbolicity is essential for a moment model, while
it is hard to obtain by a direct moment expansion of kinetic
equations. To overcome such a problem, we in \cite{framework}
fortunately developed a systematic framework to perform moment model
reduction that preserves global hyperbolicity. The framework works not
only for the models based on Hermite expansions of the distribution
function in the Boltzmann equation, but also works for any ansatz of
the distribution function in the Boltzmann equation. Actually, the
framework even works for kinetic equations in a fairly general form.

The framework developed in \cite{framework} was further presented in
the language of projection operators in \cite{Fan2015}, where the
underlying mechanism of how the hyperbolicity is preserved during the
model reduction procedure was further clarified. This is the basic
idea of our discussion in the next section.


%% file: framework.tex

\section{Theoretical Framework} \label{sec:framework}
In this section, we briefly review the framework in \cite{Fan2015} to construct globally hyperbolic
moment system from kinetic equations, as well as its variants and some further development. To
clarify the statement, we first present the definition of the hyperbolicity as follows:
\begin{definition}
  The first-order system of equations 
  \begin{equation*}
    \pd{\bw}{t} + \sum_{d=1}^D\bA_d(\bw)\pd{\bw}{x_d} = 0,
    \quad \bw \in \bbG
  \end{equation*}
  is hyperbolic at $\bw_0$, if for any unit vector $\bn\in\bbR^D$, the matrix
  $\sum_{d=1}^Dn_d\bA_d(\bw_0)$ is real diagonalizable; the system is called globally hyperbolic if
  it is hyperbolic for any $\bw\in\bbG$.
\end{definition}
Based on this definition, the analysis of the hyperbolicity of moment systems reduces to a problem
of linear algebra: the analysis of the real diagonalizablity of the coefficient matrices. Without
knowing the exact values of the matrix entries, the real diagonalizability of a matrix has to be
studied by some sufficient conditions. Some of them are
\begin{approach}\label{eigenvectors}
  all its eigenvalues are real and it has $n$ linearly independent eigenvectors.
\end{approach}
\begin{approach}\label{eigenvalues}
  all the eigenvalues of the matrix are real and distinct.
\end{approach}
\begin{approach}\label{symmetry}
  the matrix is symmetric or similar to a symmetric matrix.
\end{approach}

Grad \cite{Grad} investigated the characteristic structure of the 1D reduction of Grad's 13-moment
system, whose hyperbolicity was further studied in \cite{Muller} based on the \cref{eigenvalues}.
Afterwards, this condition is adopted in the proof of the hyperbolicity of the regularized moment
system for the 1D case in \cite{Fan}. It is worth noting that using \cref{eigenvalues} usually
requires us to compute the characteristic polynomial of the coefficient matrix of the moment system,
and for large moment systems, this may be complicated or even impractical. Even if the
characteristic polynomial is computed, showing that the eigenvalues are real and distinct is still
highly nontrivial. This severely restricts the use of this condition in kinetic model reduction.

To study the hyperbolicity in multi-dimensional cases, we have applied \cref{eigenvectors} in
\cite{Grad13toR13} to show that Grad's 13-moment system loses hyperbolicity even in an arbitrarily
small neighborhood
of the equilibrium, and in \cite{Fan_new} to prove the global hyperbolicity of the regularized moment
system for the multi-dimensional case. Due to the requirement on the eigenvectors, both proofs based
on \cref{eigenvectors} are complicated and tedious.
By contrast, it is much easier to check \cref{symmetry}, based on which Levermore provided a
concise and clear proof of the hyperbolicity of the maximum entropy moment system in
\cite{Levermore}. In \cite{Fan2015}, we re-studied the hyperbolicity of the regularized moment system
in \cite{Fan, Fan_new} based on the \cref{symmetry} and then generalized it to a framework. Below
we will start our discussion from a review of these hyperbolic moment systems.

\subsection{Review of globally hyperbolic moment system}
Let us consider the Boltzmann equation:
\begin{equation}\label{eq:boltzmann}
  \pd{f}{t} + \sum_{d=1}^D v_d \pd{f}{x_d} = Q(f),
\end{equation}
and denote the \emph{local equilibrium} by $f_{eq}$, which satisfies $Q(f_{eq})=0$ and $f_{eq}>0$.
The key idea of Grad's moment method is to expand the distribution as 
\begin{equation}\label{eq:expansion}
  f(t,\bx,\bv) = \sum_{|\alpha|\leq M}f_{eq}(t, \bx, \bv) f_{\alpha}(t, \bx)\He_{\alpha}(t,\bx,\bv)
  = \sum_{|\alpha|\leq M}f_{\alpha}(t,\bx)\mH_{\alpha}(t,\bx,\bv)
\end{equation}
for a given integer $M\geq 2$,
where for the multi-dimensional index $\alpha\in\bbN^D$, $|\alpha|=\sum_{d=1}^D\alpha_d$, and the
basis function $\mH_{\alpha}$ is defined by $\mH_{\alpha} = f_{eq}\He_{\alpha}$, with $\He_{\alpha}$
being the orthonormal polynomials of $\bv$ with weight function $f_{eq}$. When $f_{eq}$ is the local
Maxwellian, $\He_{\alpha}$ can be obtained by translation and scaling of Hermite polynomials.
Grad's moment system can then be obtained by substituting the expansion into the Boltzmann equation
and matching the coefficients of $\mH_{\alpha}$ with $|\alpha|\leq M$.
To clearly describe this procedure, we assume that the distribution function $f$ is defined on a
space $\bbH$ spanned by the basis functions $\mH_{\alpha}$ for all $\alpha \in \bbN^D$, and we let
$\bbH_M := \mathrm{span}\{\mH_{\alpha}: |\alpha|\leq M\}$ be the subspace for our model reduction.
Then one can introduce the projection from $\bbH$ to $\bbH_M$ as 
\begin{equation}\label{eq:mP}
  \mP f = \sum_{|\alpha|\leq M}f_{\alpha}\mH_{\alpha} \text{ with }
  f_{\alpha} = \langle f, \mH_{\alpha}\rangle,
\end{equation}
where the inner product is defined as $\langle f, g\rangle = \int_{\bbR^D}fg / f_{eq}\dd\bv$.
The projection accurately describes Grad's expansion \eqref{eq:expansion} and 
provides a tool to study the operators in the space $\bbH_M$. For example, matching the coefficients
of the basis $\mH_{\alpha}$ with $|\alpha|\leq M$ can be understood as projecting the system into
the space $\bbH_M$. Hence, Grad's moment system is written as
\begin{equation}\label{eq:ms_operator}
  \mP\pd{\mP f}{t} + \sum_{d=1}^D\mP v_d\pd{\mP f}{x_d} = \mP Q(\mP f).
\end{equation}

Let $\bmH$ be the vector whose components are all the basis functions $\mH_{\alpha}$ with
$|\alpha|\leq M$ listed in a given order.
Since $\mP f$ is a function in $\bbH_M$, one can collect all the independent variables in
$\mP f$ and denote it by $\bw$ with its length equal to the dimension of $\bbH_M$. Thanks to the
definition of the projection operator $\mP$, there exist the square matrices $\bD$ and $\bB_d$,
$d=1,\dots,D$ such that 
\begin{equation}\label{eq:operator2matrix}
  \mP\pd{\mP f}{t} = \bmH^T\bD\pd{\bw}{t},\quad
  \mP v_d\pd{\mP f}{x_d} = \bmH^T\bB_d\pd{\bw}{x_d}.
\end{equation}
Accordingly, letting $\bQ$ be the vector such that $\mP Q(\mP f)=\bmH^T\bQ$, one can rewrite Grad's
moment system as
\begin{equation}\label{eq:ms_linear}
  \bD\pd{\bw}{t} + \sum_{d=1}^D\bB_d\pd{\bw}{x_d} = \bQ.
\end{equation}
Actually, the system \eqref{eq:ms_linear} is the vector form of \eqref{eq:ms_operator} in $\bbH_M$
with the basis $\mH_{\alpha}$. By comparing these equations, we have the following correspondences
\begin{equation}\label{eq:correspondence}
    \bw\leftrightarrow \mP f,\quad 
    \bD\pd{}{t} \leftrightarrow \mP\pd{}{t},\quad 
    \bB_d\pd{}{x_d}\leftrightarrow \mP v_d\pd{}{x_d},\quad
    \bQ\leftrightarrow \mP Q (\mP f).
\end{equation}

\begin{figure}[ht]
  \centering
  \subfloat[Grad's moment system\label{fig:grad}]{
    \includegraphics[width=0.8\textwidth]{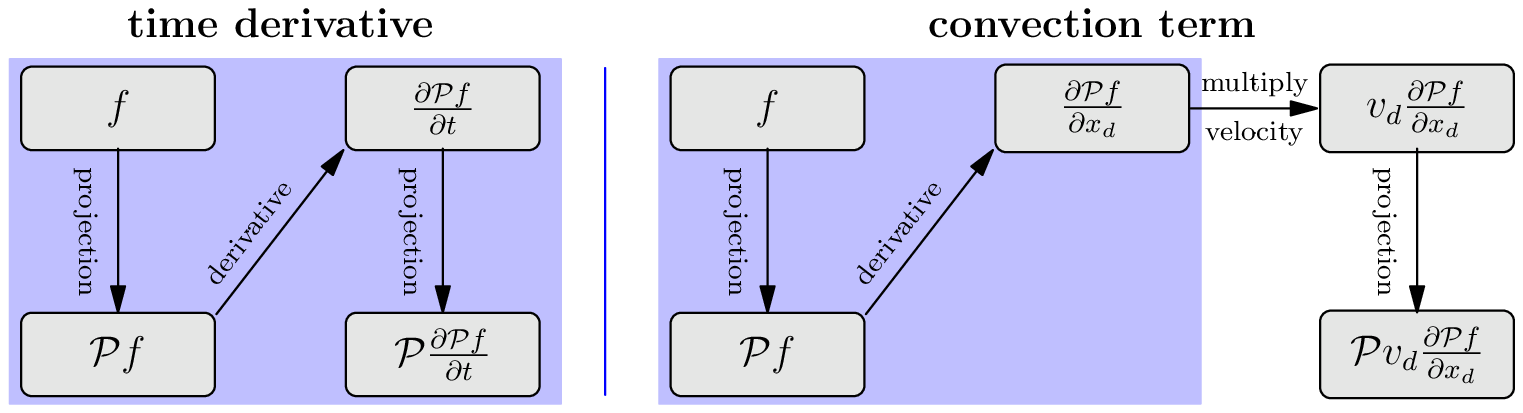}
  } \\
  \subfloat[Hyperbolic regularized moment system\label{fig:hme}]{
    \includegraphics[width=0.8\textwidth]{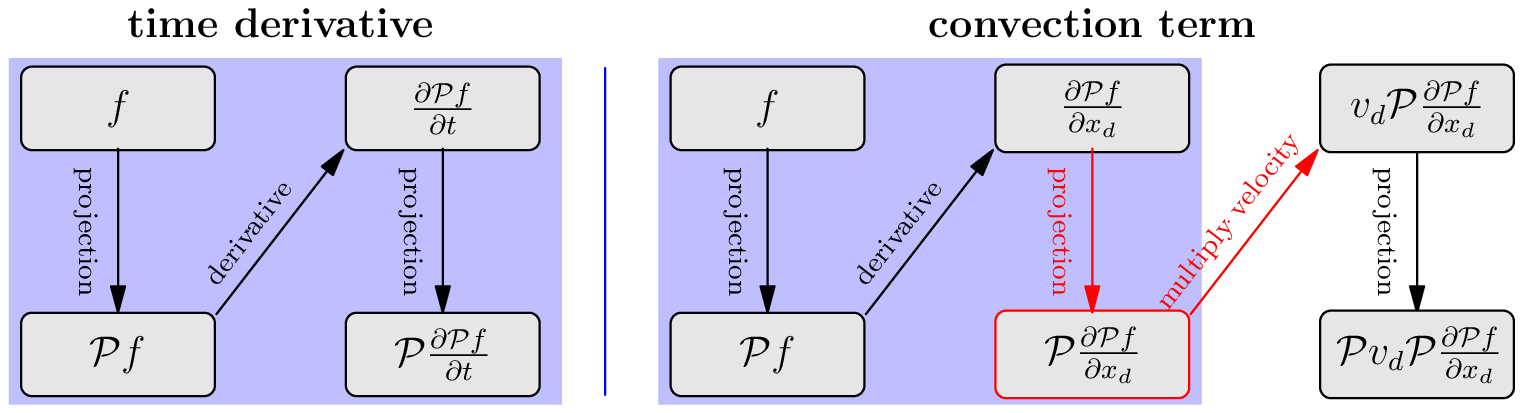}
  }
  \caption{\label{fig:diagram} Diagram of the procedure of Grad's and regularized moment system.}
\end{figure}

Furthermore, we can diagram the procedure to derive Grad's moment system in \cref{fig:grad}. It is
noticed in \cite{Fan2015} that the time derivative and the spatial derivative are treated
differently in such a process, as a projection operator is applied directly to the time derivative,
while for the spatial derivative, this projection operator appears only after the velocity $v$ is
multiplied. This difference causes the loss of hyperbolicity. By such observation, we have drawn a
key conclusion in \cite{Fan2015} that one should add a projection operator right in front of the
spatial derivative to regain hyperbolicity, as is illustrated in \cref{fig:hme}. The corresponding
moment system is 
\begin{equation}\label{eq:regularized_ms}
    \mP\pd{\mP f}{t} + \sum_{d=1}^D\mP v_d{\color{red}\mP} \pd{\mP f}{x_d} = \mP Q(\mP f),
\end{equation}
where the additional projection operator is labeled in red. Using \eqref{eq:operator2matrix}, one
can claim that there exist the square matrices $\bM_d$, $d=1,\dots,D$ such that 
\begin{equation}
  \mP v_d\mP\pd{\mP f}{x_d} = \bmH^T\bM_d\bD\pd{\bw}{x_d},
\end{equation}
and obtain the vector form of the regularized moment system as 
\begin{equation}\label{eq:regularized_ms_linear}
  \bD\pd{\bw}{t} + \sum_{d=1}^D\bM_d\bD\pd{\bw}{x_d} = \bQ.
\end{equation}
Similar to \eqref{eq:correspondence}, we have one more correspondence:
\begin{equation}
    \bM_d\leftrightarrow \mP v_d,
\end{equation}
that is to say, the matrices $\bM_d$ are the representation of the operators $\mP v_d$ on $\bbH_M$.
It is not difficult to check that the matrices $\bM_d$ are symmetric due to the orthonormality of
the basis $\mH_{\alpha}$, so that any linear combination of the matrices $\bM_d$ is real
diagonalizable. One can also check the matrix $\bD$ is invertible. Hence $\bD^{-1}\bM_d\bD$ is
similar to $\bM_d$ so that the system \eqref{eq:regularized_ms_linear} is globally hyperbolic.
Moreover, if one multiplies $\bD^T$ on both sides of \eqref{eq:regularized_ms_linear}, the resulting
system 
\begin{equation}
  \bD^T\bD\pd{\bw}{t} + \sum_{d=1}^D\bD^T\bM_d\bD\pd{\bw}{x_d} = \bD^T\bQ
\end{equation}
turns out to be a symmetric hyperbolic system of balance laws.

\subsection{Hyperbolic regularization framework}
Till now, the hyperbolicity of \eqref{eq:regularized_ms_linear} has been proved using the
\cref{symmetry}. Looking back on the whole procedure, one can find that the key point of
the hyperbolic regularization is the extra projection operator in front of the spatial
differentiation operator in \eqref{eq:regularized_ms}. Meanwhile, the underlying mechanism to obtain
hyperbolicity can be extended to much more general cases. For example, the radiative transfer
equation has the form
\begin{multline*}
  \pd{f(t, \bx, \theta, \varphi)}{t} + \bxi(\theta,\varphi) \cdot \nabla_{\bx}
  f(t,\bx,\theta,\varphi) = Q(f)(t,\bx,\theta,\varphi), \\
  \bx \in \mathbb{R}^3, \quad \theta \in [0,\pi), \quad \varphi \in [0,2\pi),
\end{multline*}
where the velocity is given by $\bxi(\theta,\varphi) = (\sin\theta \cos\varphi, \sin\theta
\sin\varphi, \cos\theta)^T$. To derive reduced models, one can replace the local equilibrium
$f_{eq}$ in \eqref{eq:expansion} by a nonnegative weight function $\omega$, and correspondingly,
the orthogonal polynomials $\He_{\alpha}$ should be replaced by the orthogonal basis functions
$\phi_{\alpha}$ for the $L^2$ space weighted by $\omega$, so that the basis functions $\mH_{\alpha}$
become $\Phi_{\alpha}:=\omega\phi_{\alpha}$. By letting $\bbH_M :=
\mathrm{span}\{\Phi_{\alpha}: |\alpha|\leq M\}$, one can similarly define the projection operator
$\mP$ as in \eqref{eq:mP}. As an extension of the globally hyperbolic moment system, we obtain 
\begin{equation}\label{eq:framework}
  \mP\pd{\mP f}{t} + \sum_{d=1}^D\mP\xi_d(\theta,\varphi)\mP\pd{\mP f}{x_d} = \mP Q(\mP f).
\end{equation}
Again, if the corresponding matrix $\bD$ as in \eqref{eq:ms_linear} is invertible, the resulting
moment system is globally hyperbolic. We refer the readers to \cite{framework, Fan2015, HMPn} for
more details of such applications in radiative transfer equations.

This framework provides a concise and clear procedure to derive the hyperbolic moment system from
a broad range of kinetic equations. It has been applied to many fields, including anisotropic
hyperbolic moment system for Boltzmann equation \cite{ANRxx}, semiconductor device simulation
\cite{Tiao}, plasma simulation \cite{Wang}, density functional theory \cite{DFT}, quantum gas
kinetic theory \cite{di2016quantum}, and rarefied relativistic Boltzmann equation
\cite{kuang2017globally}. 

\subsection{Further progress}
The above framework provides an approach to handling the hyperbolicity of the moment system.
However, the hyperbolicity is not the only concerned property. Preserving the hyperbolicity and
other properties at the same time is often required in model reduction. Below we will list some
recent attempts in this direction.

One of the interesting properties is to recover the asymptotic limits of the kinetic equations. For
example, the first-order asymptotic hydrodynamic limit of the Boltzmann equation is the
Navier-Stokes equations, and therefore it is desirable that the moment equations can preserve such a
limit. For the classical Boltzmann equation, most moment systems can automatically preserve the
Navier-Stokes limit if the stress tensor and heat flux are included.
However, for the quantum Boltzmann equation, the equilibrium has a very special form, so that the
moment system directly derived from the framework by taking the equilibrium as the weight function
disobeys the Navier-Stokes limit \cite{di2016quantum}. In this case, the authors of
\cite{di2016quantum} proposed a method called \emph{local linearization} to regularize the moment
system. Specifically, we assume the Grad-type system has the form as \eqref{eq:ms_linear} and define
$\hat{\bM}_d(\bw)=\bB_d(\bw)\bD(\bw)^{-1}$. In the regularization, the matrix $\hat{\bM}_d(\bw)$ is
replaced by $\bM_d:=\hat{\bM}_d(\bw_{eq})$ with $\bw_{eq}$ being the local equilibrium of the state
$\bw$. Such a method allows us to acquire both the hyperbolicity and Navier-Stokes limit
simultaneously. The symmetry of $\bM$ is thereby lost so that one has to use \cref{eigenvectors} to
prove the hyperbolicity.

Another relevant work is the nonlinear moment system for radiative transfer equation in \cite{MPn,
HMPn}. In order to retain the diffusion limit (similar to the Navier-Stokes limit for the Boltzmann
equation), the authors pointed out that the projection operators in \eqref{eq:framework} at
different places do not have to be same and revised \eqref{eq:framework} to be
\begin{equation}
  \tilde{\mP}\pd{\mP f}{t} + \sum_{d=1}^D\tilde{\mP}\xi_d(\theta,\varphi)\tilde{\mP}\pd{\mP f}{x_d}
  = \tilde{\mP} Q(\mP f).
\end{equation}
The operators $\mP$ and $\tilde{\mP}$ are orthogonal projections onto different subspaces of $\bbH$.
By a careful choice of the subspace for the operator $\tilde{\mP}$, the diffusion limit can be
achieved, and meanwhile, the symmetry of $\bM$ corresponding to that in
\eqref{eq:regularized_ms_linear} is preserved, leading again to global hyperbolicity.  This
generalization has broadened the application the hyperbolic regularization framework and also
permits us to take more properties of the kinetic equations into account.

Besides the hyperbolicity for the convection term, one may also be interested in the wellposedness
of the complete moment system including the collision term. One related property is Yong's first
stability condition \cite{yong1999singular}, which includes the constraints on the convection term,
collision term, and the coupling of both. This stability condition is shown to be critical for
the existence of the solutions in \cite{YJPeng2016}. In \cite{LinearStabilityHME}, the authors have
studied multiple Grad-type moment systems and confirmed that all of these systems satisfy Yong's
first stability condition.

Under this concise and flexible framework, one may wonder what is sacrificed for the hyperbolicity.
By writing out the equations, one can immediately observe that the form of balance law is ruined by
the hyperbolic regularization. A natural question is: how to define the discontinuity in the
solution? More generally, one may ask: what is the effect of such a regularization on the accuracy
of the model? In the following section, we will provide some clues using numerical experiments.



%% file: latest.tex
\section{Numerical Validation}
The application of the framework in the gas kinetic theory has been
investigated in a number of works \cite{Fan, Qiao, Microflows1D, Cai2018},
where many one- and two-dimensional examples have been numerically
studied to show the validity of hyperbolic moment equations. However, these
globally hyperbolic models, as an improvement of Grad's original models, have
never been compared with Grad's models in terms of the modeling accuracy. The
only direct comparison seen in the literature is in \cite{Microflows1D}, wherein
for a shock tube problem with a density ratio of $7.0$, the simulation of Grad's
moment equations breaks down and the corresponding hyperbolic moment equations
appear to be stable. Without running numerical tests for the same problem for
which both models work and comparing the results, it could be questioned
whether we lose accuracy when fixing the hyperbolicity. Such doubt may arise
since the globally hyperbolic models can be considered as a partial
linearization of Grad's models about the local Maxwellians.

In this section, we will make such straightforward comparison using the same
numerical examples for both methods. For simplicity, we only consider the
one-dimensional physics, for which both $x$ and $v$ are scalars. In this case,
the characteristic polynomial for the Jacobian of the flux function has an
explicit formula \cite{Fan}, so that the hyperbolicity of Grad's equation can
be easily checked. The underlying kinetic equation used in our test is the
Boltzmann-BGK equation with a constant relaxation time
\begin{equation} \label{eq:BGK}
\frac{\partial f}{\partial t} + v \frac{\partial f}{\partial x}
  = \frac{1}{\mathit{Kn}} (f_{eq} - f).
\end{equation}
The ansatz of the distribution function is given by \eqref{eq:mP}, so that
\eqref{eq:ms_operator} stands for Grad's moment system, and
\eqref{eq:regularized_ms} stands for the hyperbolic moment system. Below we
are going to use two benchmark tests to show the performance of both types of
models. In general, both Grad's moment equations and the hyperbolic moment
equations are solved by the first-order finite volume method with local
Lax-Friedrichs numerical flux. Time splitting is applied to solve the advection
part and the collision part separately, and for each part, the forward Euler
method is applied. The CFL condition is utilized to determine the time step,
and the Courant number is chosen as $0.9$. For Grad's moment method, the
maximum characteristic speed is obtained by solving the roots of the
characteristic polynomial of the Jacobian, and the explicit expression of the
charateristic polynomial has been given in \cite{Fan}. For the hyperbolic
moment method, the maximum characteristic speeds have been computed in
\cite{Fan}. The explicit form of the hyperbolic moment system (given in
\cite{Fan}) shows that its last equation contains a non-conservative product,
which is discretized by central difference. In all the numerical examples, the
number of grid cells is $1000$ if not otherwise specified. We have done the
convergence test showing that for smooth solutions, such a resolution can
provide solutions sufficiently close to the solutions on a much finer grid, so
that their difference is invisible to the naked eye. When exhibiting the
numerical results, we will mainly focus on the equilibrium variables including
density $\rho$, velocity $u$, and temperature $\theta$, which are defined by
\begin{align*}
\rho(t,x) &= \int_{\mathbb{R}} f(t,x,v) \,\mathrm{d}v, \\
u(t,x) &= \frac{1}{\rho(t,x)} \int_{\mathbb{R}} v f(t,x,v) \,\mathrm{d}v, \\
\theta(t,x) &= \frac{1}{\rho(t,x)} \int_{\mathbb{R}} [v-u(t,x)]^2 f(t,x,v) \,\mathrm{d}v.
\end{align*}

\subsection{Shock structure}
The structure of plane shock waves is frequently used as a benchmark test in
the gas kinetic theory. It shows that the physical shock, which appears to be a
discontinuity in the Euler equations, is actually a smooth transition from one
state to another. The computational domain is $(-\infty, +\infty)$ so that no
boundary condition is involved, and the initial data are
\begin{equation} \label{eq:init_shock}
f(0,x,v) = \left\{ \begin{array}{ll}
  \dfrac{\rho_l}{\sqrt{2\pi \theta_l}}
    \exp \left( -\dfrac{(v - u_l)^2}{2\theta_l} \right), & \text{if } x < 0, \\[10pt]
  \dfrac{\rho_r}{\sqrt{2\pi \theta_r}}
    \exp \left( -\dfrac{(v - u_r)^2}{2\theta_r} \right), & \text{if } x > 0,
\end{array} \right.
\end{equation}
where all the equilibrium variables are determined by the Mach number
$\mathit{Ma}$:
\begin{gather*}
\rho_l = 1, \qquad u_l = \sqrt{3} Ma, \qquad \theta_l = 1, \\
\rho_r = \frac{2 \mathit{Ma}^2}{\mathit{Ma}^2 + 1}, \qquad
u_r = \frac{\sqrt{3} \mathit{Ma}}{\rho_r}, \qquad
\theta_r = \frac{3\mathit{Ma}^2 - 1}{2\rho_r}.
\end{gather*}
We are interested in the steady-state of this problem. Since the parameter
$\mathit{Kn}$ only introduces a uniform spatial scaling, it does not affect the
shock structure. Therefore we simply set it to be $1$. Numerically, we set the
computational domain to be $[-30, 30]$. The boundary condition is provided by
the ghost-cell method, and the distribution functions on the ghost cells are
set to be the two states defined in \eqref{eq:init_shock}.

\subsubsection{Case 1: $\mathit{Ma} = 1.4$ and $M = 4$}
\begin{figure}[!ht]
\centering
\subfloat[Grad vs HME\label{fig:Ma_1p4_comp}]{\includegraphics[scale=.48]{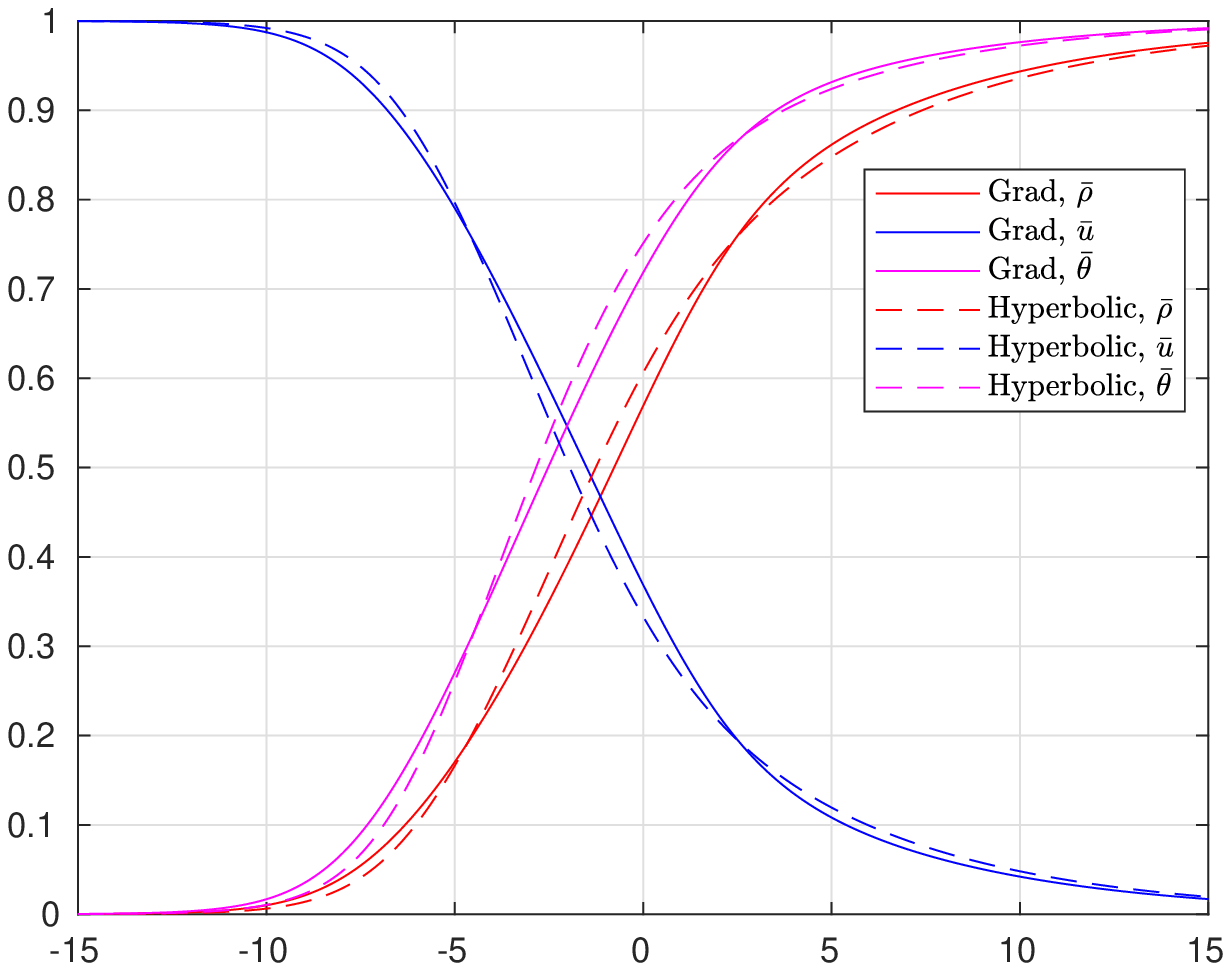}}
\hspace{10pt}
\subfloat[Phase diagram of Grad's solution\label{fig:Ma_1p4_phase}]{\includegraphics[scale=.48]{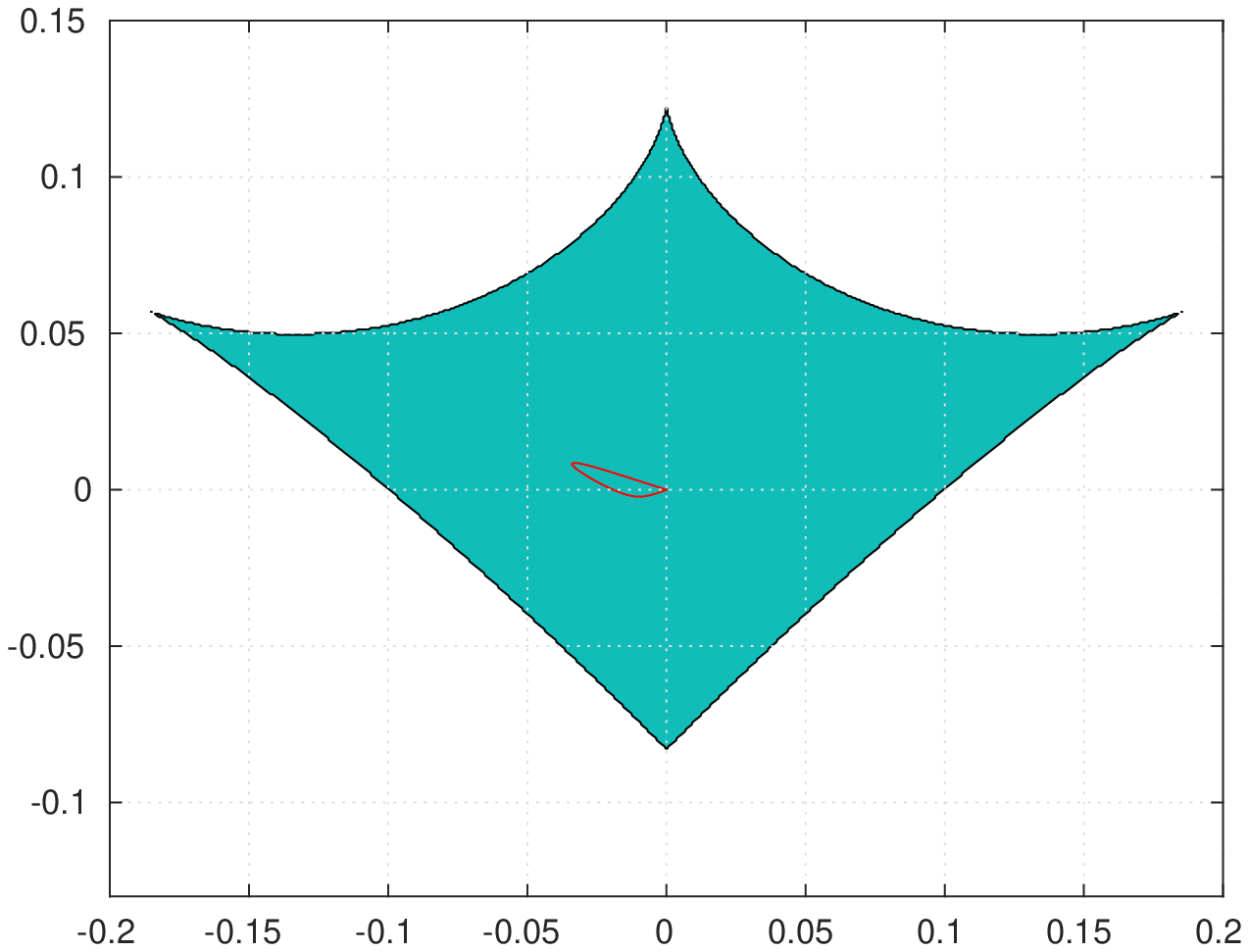}}
\caption{Left: The comparison of shock structures of two solutions with Mach
number $1.4$ and $M = 4$. Right: The green area is the hyperbolicity region
(horizontal axis: $\hat{f}_{M-1}$, vertical axis: $\hat{f}_M$), and the red
loop is the parametric curve $(\hat{f}_{M-1}, \hat{f}_M)$ with parameter $x$.}
\label{fig:Ma_1p4}
\end{figure}

In this case, both Grad's system and the hyperbolic moment system work due to
the relatively small Mach number. The numerical results are shown in Figure
\ref{fig:Ma_1p4}. By convention, we plot the normalized density, velocity, and
temperature defined by
\begin{displaymath}
\bar{\rho}(x) = \frac{\rho(x) - \rho_l}{\rho_r - \rho_l}, \quad
\bar{u}(x) = \frac{u(x) - u_r}{u_l - u_r}, \quad
\bar{\theta}(x) = \frac{\theta(x) - \theta_l}{\theta_r - \theta_l},
\end{displaymath}
so that the value of all variables are generally within the range $[0,1]$,
unless the temperature overshoot is observed.

Figure \ref{fig:Ma_1p4_phase} shows the hyperbolicity region of Grad's moment
equations. It has been proven in \cite{Fan} that for the one-dimensional
physics, the hyperbolicity region can be characterized by the following two
dimensionless quantities:
\begin{displaymath}
\hat{f}_{M-1} = \frac{f_{M-1}}{\rho \theta^{(M-1)/2}}, \qquad
\hat{f}_M = \frac{f_M}{\rho \theta^{M/2}},
\end{displaymath}
where $f_M$ and $f_{M-1}$ are the last two coefficients in the expansion
\eqref{eq:mP}. The red curve in Figure \ref{fig:Ma_1p4_phase} provides the
trajectory of Grad's solution in this diagram. It can be seen that for such a
small Mach number, the whole solution is well inside the hyperbolicity region,
so that the simulation of Grad's moment equations is stable. Figure
\ref{fig:Ma_1p4_comp} shows that both methods provide smooth shock structures,
and the predictions for all the equilibrium variables are similar. This example
confirms the applicability of both systems in weakly non-equilibrium regimes.
Note that for one-dimensional physics, Grad's equations do not suffer form the 
loss of hyperbolicity near equilibrium.

\subsubsection{Case 2: $\mathit{Ma} = 2.0$ and $M = 4$}
\begin{figure}[!ht]
\centering
\subfloat[Grad vs HME\label{fig:Ma_2_comp}]{\includegraphics[scale=.48]{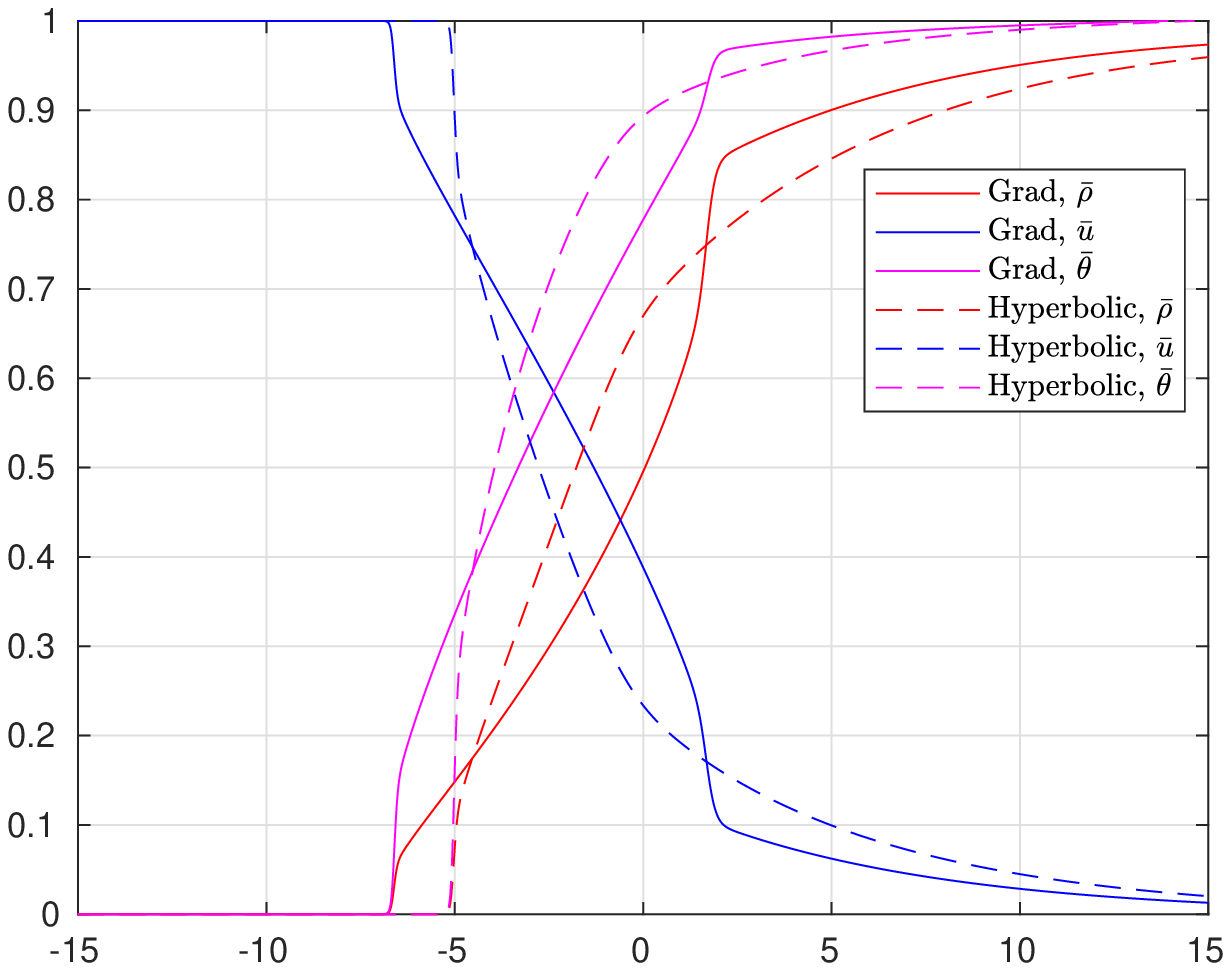}}
\hspace{10pt}
\subfloat[Phase diagram of Grad's solution\label{fig:Ma_2_phase}]{\includegraphics[scale=.48]{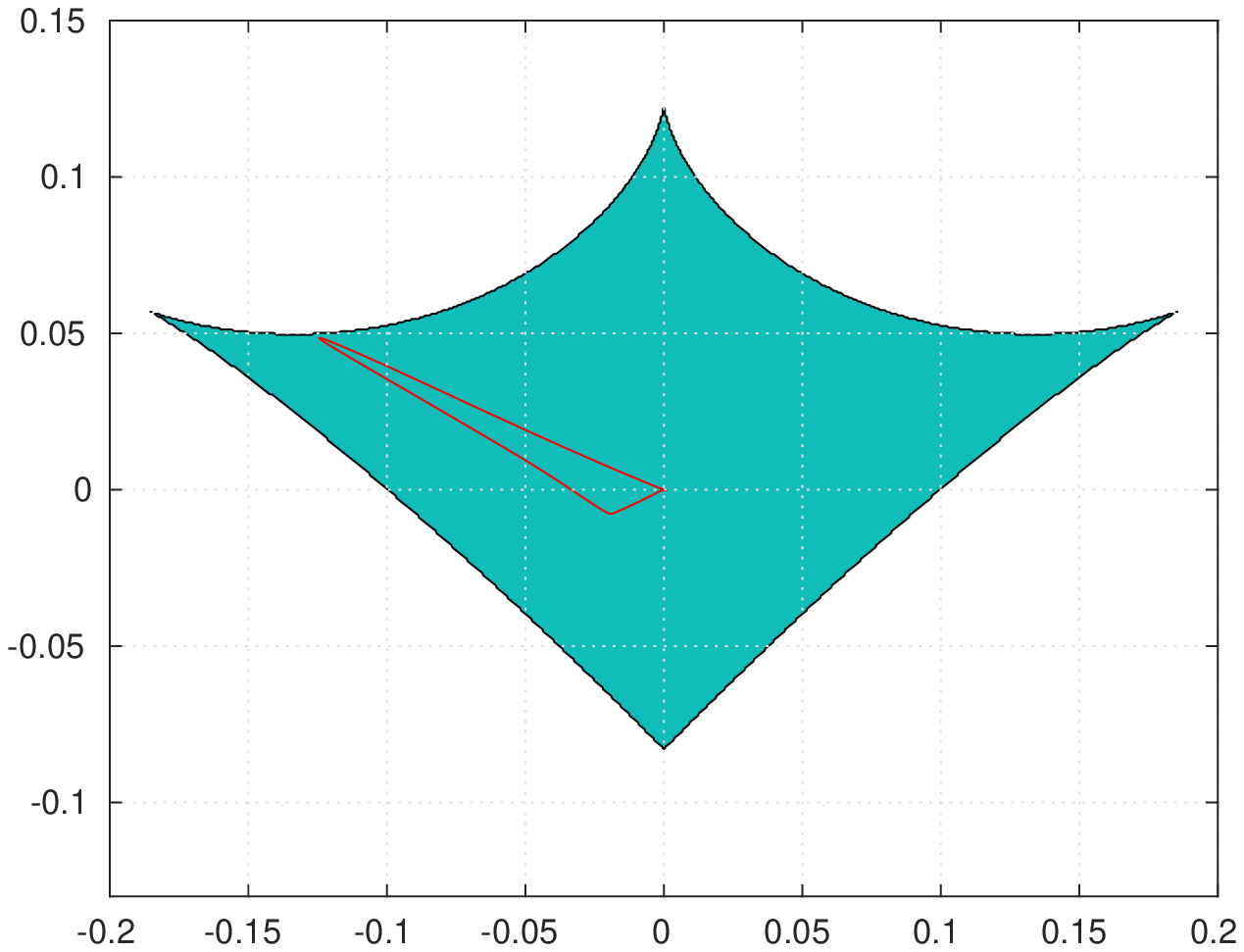}}
\caption{Left: The comparison of shock structures of two solutions with Mach
number $2.0$ and $M = 4$. Right: The green area is the hyperbolicity region
(horizontal axis: $\hat{f}_{M-1}$, vertical axis: $\hat{f}_M$), and the red
loop is the parametric curve $(\hat{f}_{M-1}, \hat{f}_M)$ with parameter $x$.}
\label{fig:Ma_2}
\end{figure}

Now we increase the Mach number to introduce stronger non-equilibrium. The same
plots are provided in Figure \ref{fig:Ma_2}. In this example, despite the
numerical diffusion, discontinuities can be identified without difficulty from
the numerical solutions. These discontinuities, also known as subshocks, appear
due to the insufficient characteristic speed in front of the shock wave,
meaning that both systems are insufficient to describe the physics. To capture
these discontinuities, $8000$ grid cells are used in the spatial discretization.
This example shows significantly different shock structures predicted by both
methods. For Grad's moment equations, the subshock locates near $x = -7$, while
for hyperbolic moment equations, the subshock appears near $x = -5$. The wave
structures also differ a lot. By focusing on the high-density region, we find
that the solution of hyperbolic moment equations is smoother, showing the
possibly better description of the physics.

Here we remind the readers that the wave structure of hyperbolic moment
equations may depend on the numerical method, due to its non-conservative
nature. The locations and the strengths of the subshock may change when using
the different shock conditions. However, we would like to argue that it is
meaningless to justify any solution with subshocks for the hyperbolic moment
equations, for it is unphysical and should not appear in the solution of the
Boltzmann equation. In practice, the appearance of discontinuous solutions is
an indication of the inadequate truncation of series, which inspires us to
increase $M$ to get more reliable solutions without subshocks.

Figure \ref{fig:Ma_2_phase} shows that Grad's solution still locates within the
hyperbolicity region, although the curve is already quite close to the boundary
of the region. This example shows that even in its hyperbolicity region, Grad's
moment method may lose its validity.

\subsubsection{Case 3: $\mathit{Ma} = 2.0$ and $M = 6$}
\begin{figure}[!ht]
\centering
\subfloat[Grad vs HME\label{fig:Ma_2_M_6_comp}]{\includegraphics[scale=.48]{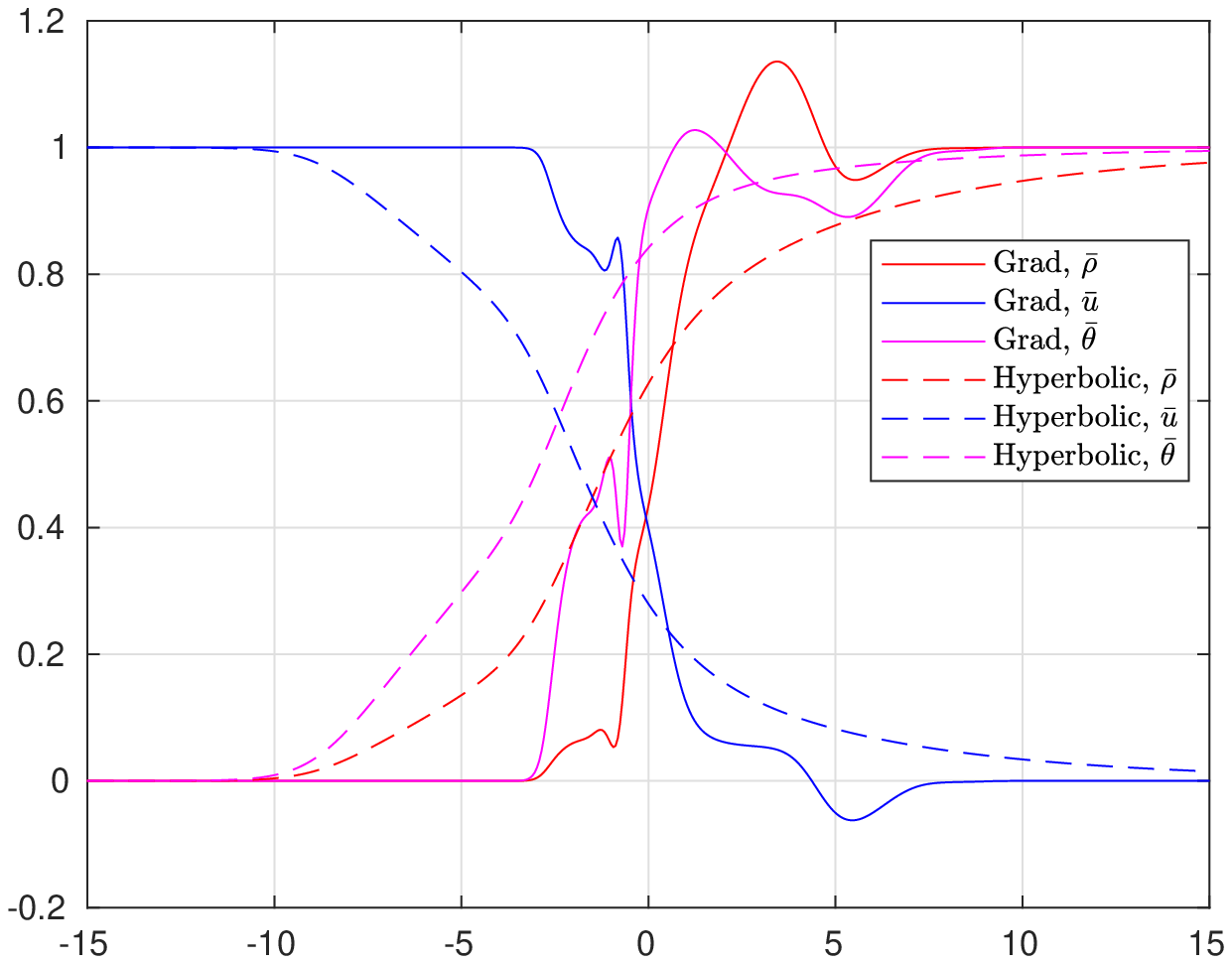}}
\hspace{9pt}
\subfloat[Phase diagram of Grad's solution\label{fig:Ma_2_M_6_phase}]{\includegraphics[scale=.48,clip]{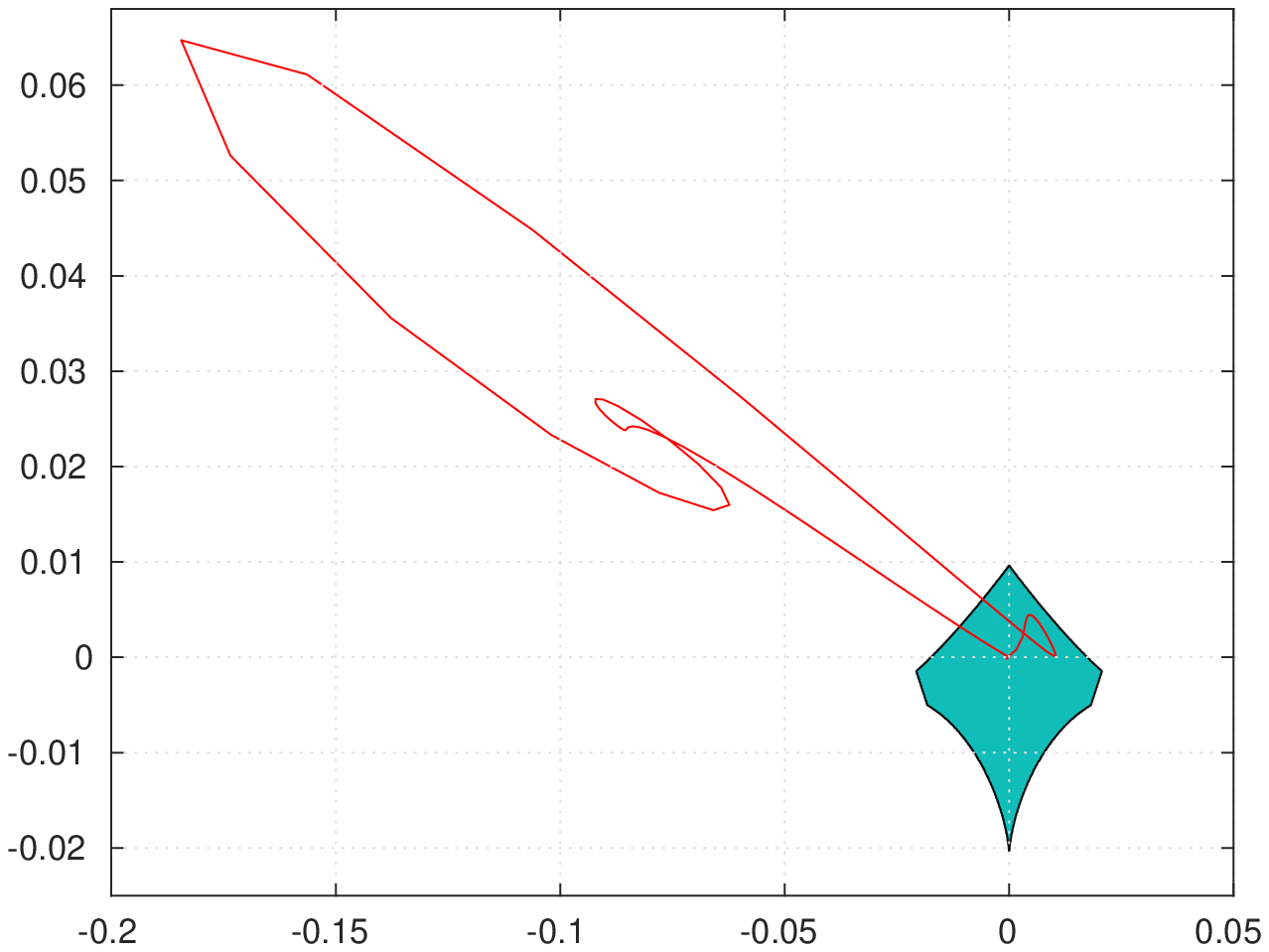}}
\caption{Left: The shock structure of hyperbolic moment equations for Mach
number $2.0$ and $M = 6$, and Grad's solution before computational failure ($t
= 1.0$).  Right: The green area is the hyperbolicity region (horizontal axis:
$\hat{f}_{M-1}$, vertical axis: $\hat{f}_M$), and the red loop is the
parametric curve $(\hat{f}_{M-1}, \hat{f}_M)$ with parameter $x$.}
\label{fig:Ma_2_M_6}
\end{figure}

Now we try to increase $M$ and carry out the simulation again for Mach number
$2.0$. The results are given in Figure \ref{fig:Ma_2_M_6}. With the hope that a
larger $M$ can provide a better solution, we actually see that Grad's moment
equations lead to computational failure. The numerical solution before the
computation breaks down is plotted in Figure \ref{fig:Ma_2_M_6_comp}. Figure
\ref{fig:Ma_2_M_6_phase} clearly shows that this is caused by the loss of
hyperbolicity. We believe that this implies the non-existence of the solution.

On the contrary, the simulation of hyperbolic moment equations is still stable.
As expected, it provides a smooth shock structure and improves the result
predicted by $M = 4$.

\subsubsection{Case 4: $\mathit{Ma} = 1.7$ and $M = 6$}
\begin{figure}[!ht]
\centering
\subfloat[Grad vs HME\label{fig:Ma_1p7_comp}]{\includegraphics[scale=.48]{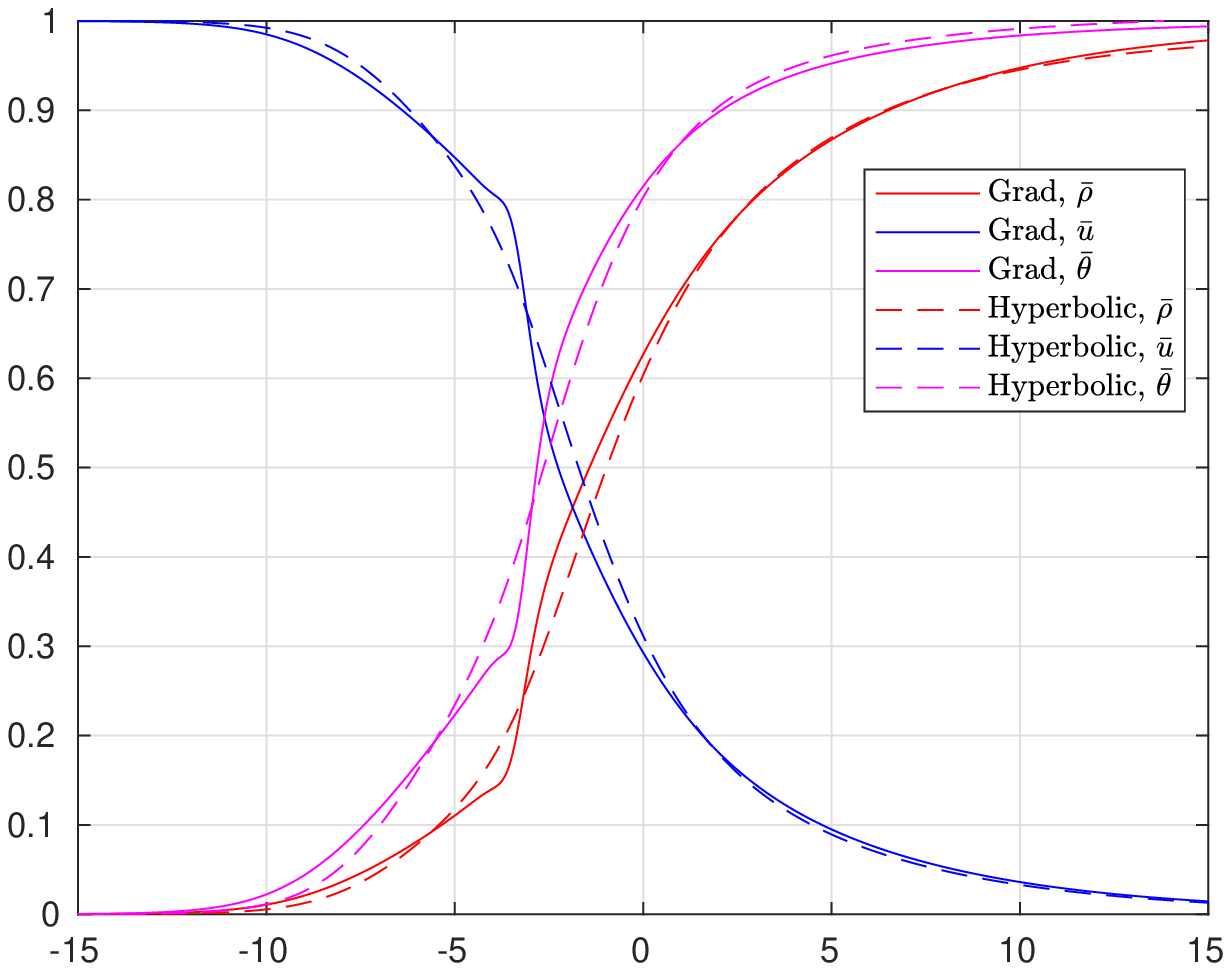}}
\hspace{10pt}
\subfloat[Phase diagram of Grad's solution\label{fig:Ma_1p7_phase}]{\includegraphics[scale=.48]{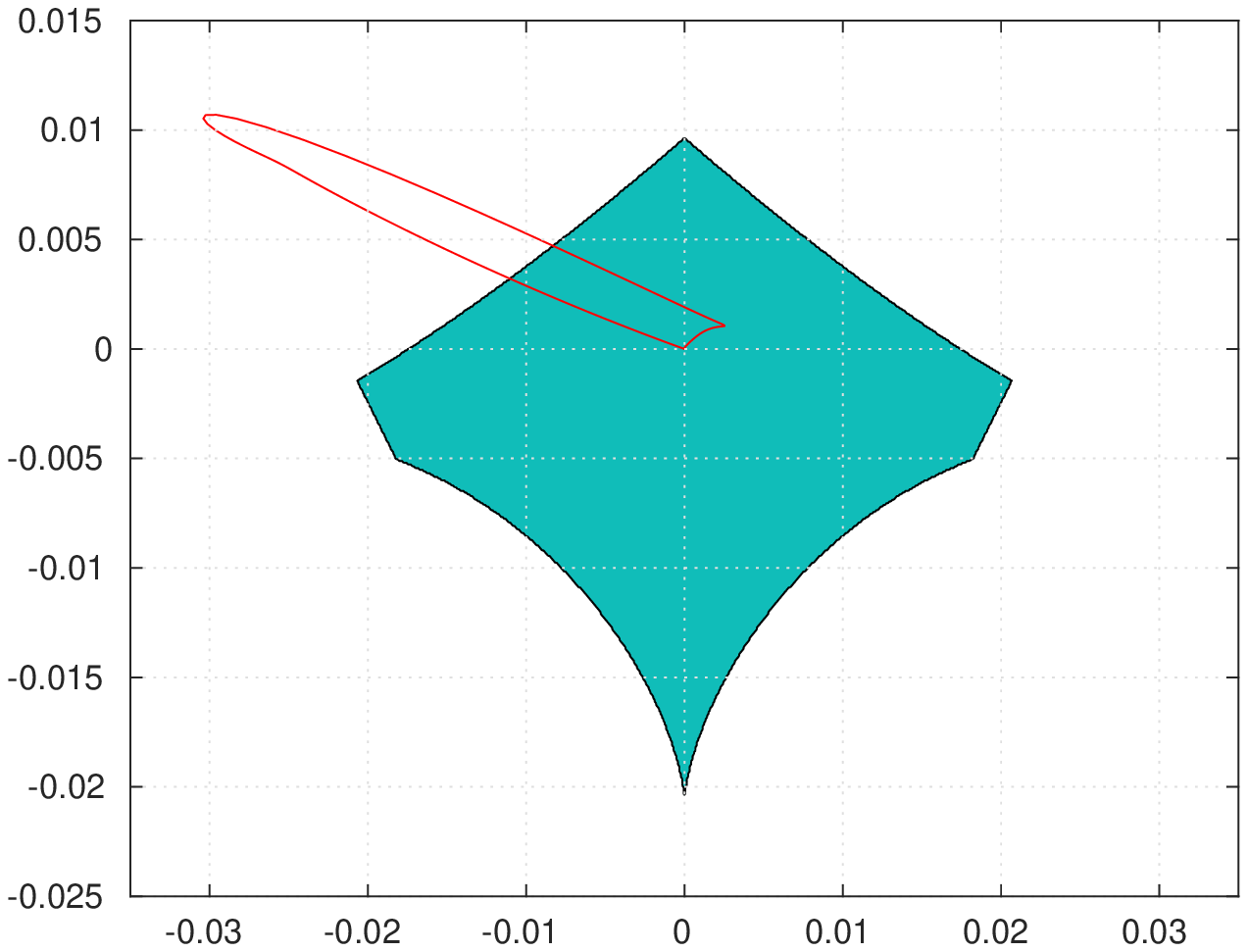}}
\caption{Left: The comparison of shock structures of two solutions with Mach
number $1.7$ and $M = 6$. Right: The green area is the hyperbolicity region
(horizontal axis: $\hat{f}_{M-1}$, vertical axis: $\hat{f}_M$), and the red
loop is the parametric curve $(\hat{f}_{M-1}, \hat{f}_M)$ with parameter $x$.}
\label{fig:Ma_1p7}
\end{figure}

In this example, we decrease the Mach number so that the shock structure of
Grad's equations can be found. Figure \ref{fig:Ma_1p7_comp} shows that the
results of both systems generally agree with each other, but it can be observed
that hyperbolic moment equations provide smoother solutions than Grad's system,
so that it is likely to be more accurate. Therefore, despite the higher
nonlinearity of Grad's system, it does not necessarily help provide better
solutions.

Interestingly, when looking at the phase diagram plotted in Figure
\ref{fig:Ma_1p7_phase}, we see that Grad's solution has run out of the
hyperbolicity region. It is to be further studied why the solution is still
stable. Here we would like to conjecture that the collision term and the
numerical diffusion help stabilize the numerical solution in the evolutionary
process, and for the steady-state equations, solutions for non-hyperbolic
equations may still exist. Nevertheless, all the above numerical tests show the
superiority of hyperbolic moment equations for both accuracy and stability.

\subsubsection{Case 5: $\mathit{Ma} = 2.0$ and $M = 10$}
\begin{figure}[!ht]
\centering
\includegraphics[scale=.48]{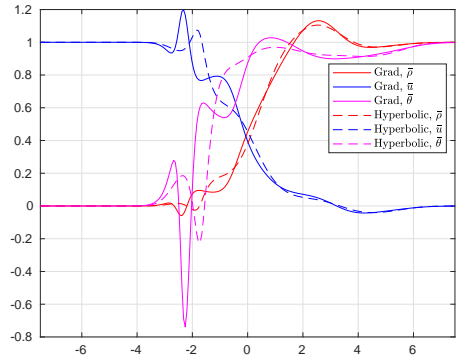}
\caption{The numerical solution at $t = 0.8$ for Mach number $2.0$ and $M = 10$.}
\label{fig:Ma_2_M_10}
\end{figure}

In this example, we would like to show the failure of both systems for a larger
$M$. In Figure \ref{fig:Ma_2_M_10}, we plot the results at $t = 0.8$, where
both numerical solutions contain negative temperatures. In \cite{Holway1965},
the reason for such a phenomenon has been explained, which lies in the
divergence of the approximation \eqref{eq:mP} as $M$ tends to infinity. It is
rigorously shown in \cite{Cai2019} that when $\theta_r > 2\theta_l$, for the
solution of the steady-state BGK equation, the limit of $\mP f$ (see
\eqref{eq:mP}) as $M \rightarrow \infty$ does not exist. Here for $\mathit{Ma}
= 2.0$, the temperature behind the shock wave is $\theta_r = 55/16 > 2 =
2\theta_l$. Thus for a large $M$, the divergence leads to a poor approximation
of the distribution function, and it is reflected as a negative temperature in
the numerical results. Such a divergence issue is independent of the subshock
and the hyperbolicity, and should be regarded as a defect for both systems. The
work on fixing the issue is ongoing.

\subsection{Fourier flow}
In this test, we are interested in the performance of both methods with wall
boundary conditions. The fluid we are concerned about is between two fully
diffusive walls locating at $x = -1/2$ and $x = 1/2$. For the Boltzmann-BGK
equation \eqref{eq:BGK}, the boundary condition is
\begin{align*}
f(t,-1/2,v) &= \frac{\rho_l}{\sqrt{2\pi\theta_l}}
  \exp\left( -\frac{v^2}{2\theta_l} \right), \qquad v > 0, \\
f(t,1/2,v) &= \frac{\rho_r}{\sqrt{2\pi\theta_r}}
  \exp\left( -\frac{v^2}{2\theta_r} \right), \qquad v < 0,
\end{align*}
where $\theta_{l,r}$ stands for the temperature of the walls, and $\rho_{l,r}$
is chosen such that
\begin{displaymath}
\int_{\mathbb{R}} v f(t,\pm 1/2,v) \,\mathrm{d}v = 0.
\end{displaymath}
Following \cite{Grad}, the boundary conditions of moment equations can be
derived by taking odd moments of the diffusive boundary condition. We choose
the initial condition as
\begin{equation} \label{eq:init}
f(0,x,v) = \frac{1}{\sqrt{2\pi}} \exp \left( -\frac{v^2}{2} \right)
\end{equation}
for all $x$. Again we are concerned only about the steady-state of the
solution.

\begin{figure}[!ht]
\centering
\subfloat[Grad vs HME\label{fig:Tr1p9_comp}]{\includegraphics[scale=.48]{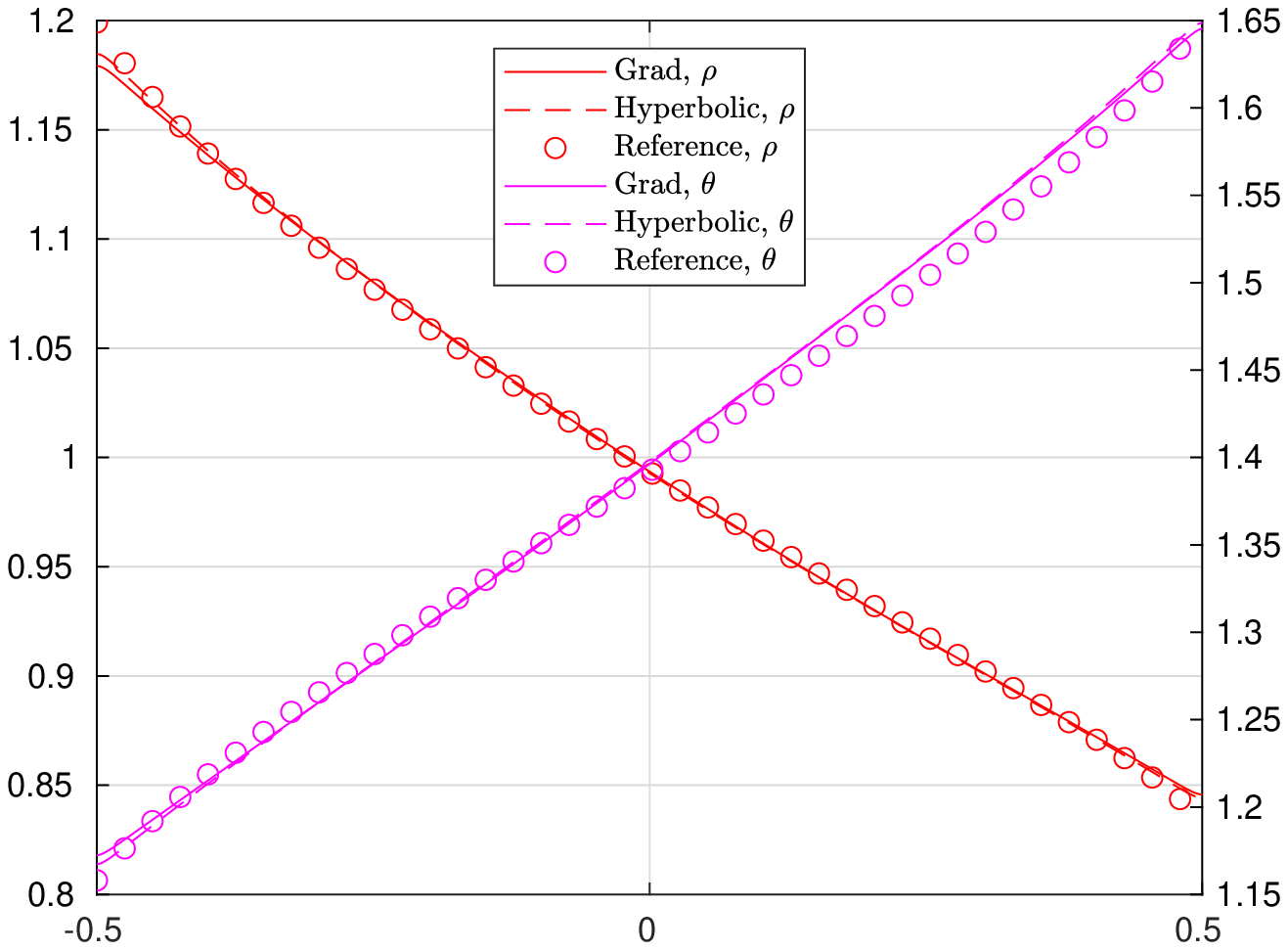}}
\hspace{10pt}
\subfloat[Phase diagram of Grad's solution\label{fig:Tr1p9_phase}]{\raisebox{-8pt}{\includegraphics[scale=.48,clip]{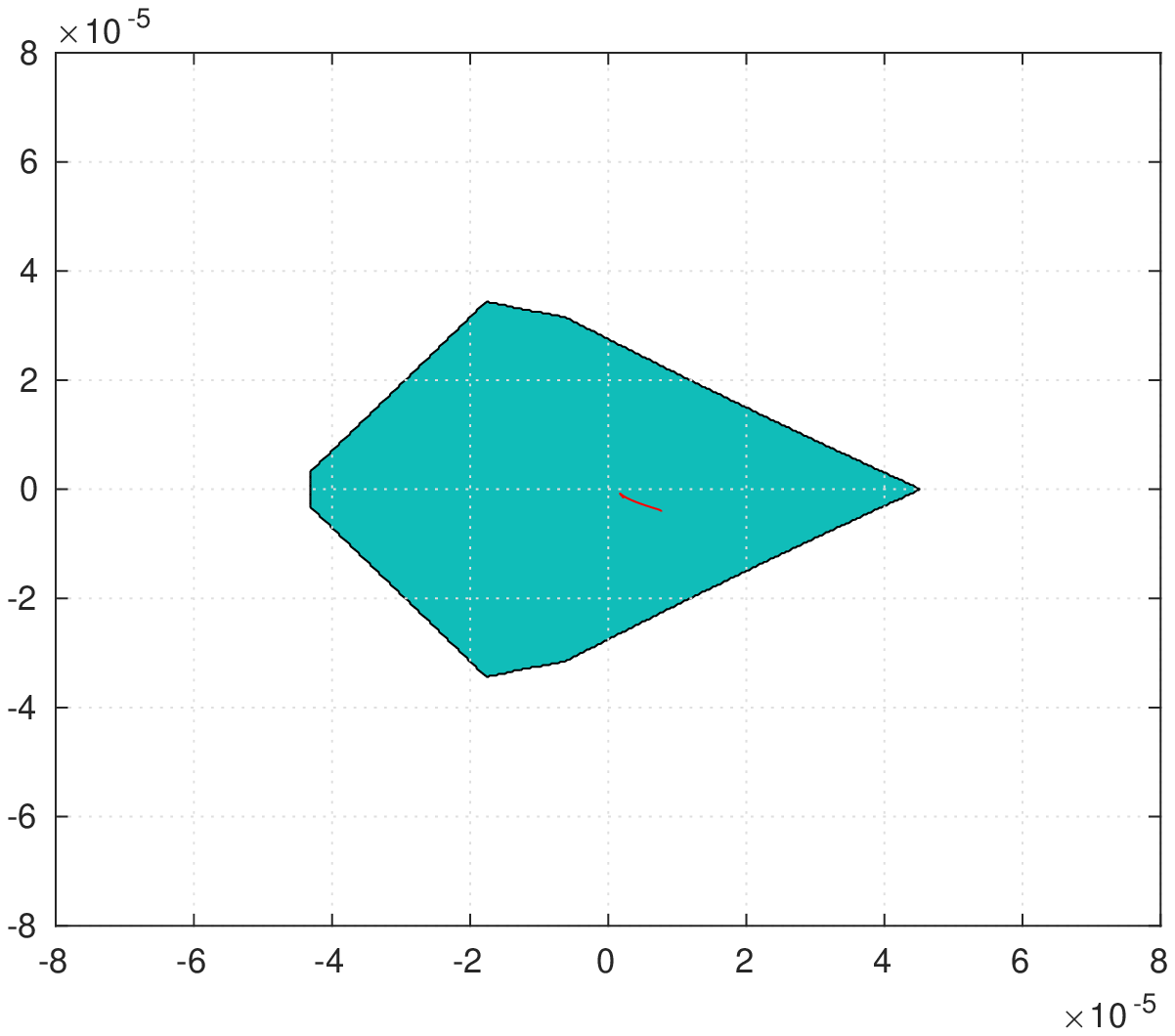}}}
\caption{Left: Steady Fourier flow for $\theta_r = 1.9$ (left vertical axis:
$\rho$, right vertical axis: $\theta$). Right: The green area is the
hyperbolicity region (horizontal axis: $\hat{f}_{M-1}$, vertical axis:
$\hat{f}_M$), and the red line is the parametric curve $(\hat{f}_{M-1},
\hat{f}_M)$ with parameter $x$.}
\label{fig:Tr1p9}
\end{figure}

In our numerical experiments, we choose $\mathit{Kn} = 0.3$, $\theta_l = 1$ and
$M = 11$. Two test cases with $\theta_r = 1.9$ and $\theta_r = 2.7$ are
considered. For the smaller temperature ratio $\theta_r = 1.9$, the numerical
results are given in Figure \ref{fig:Tr1p9}, where two solutions mostly agree
with each other.  The reference solution, computed using the discrete velocity
model, is also provided in Figure \ref{fig:Tr1p9_comp}. It can be seen that
both models provide reasonable approximations to the reference solution. The
good behavior of Grad's solutions can also be predicted by the phase diagram in
Figure \ref{fig:Tr1p9_phase}, from which one can observe that the whole
solution locates in the central area of the hyperbolicity region.

\begin{figure}[!ht]
\centering
\subfloat[Grad vs HME\label{fig:Tr2p7_comp}]{\includegraphics[scale=.48]{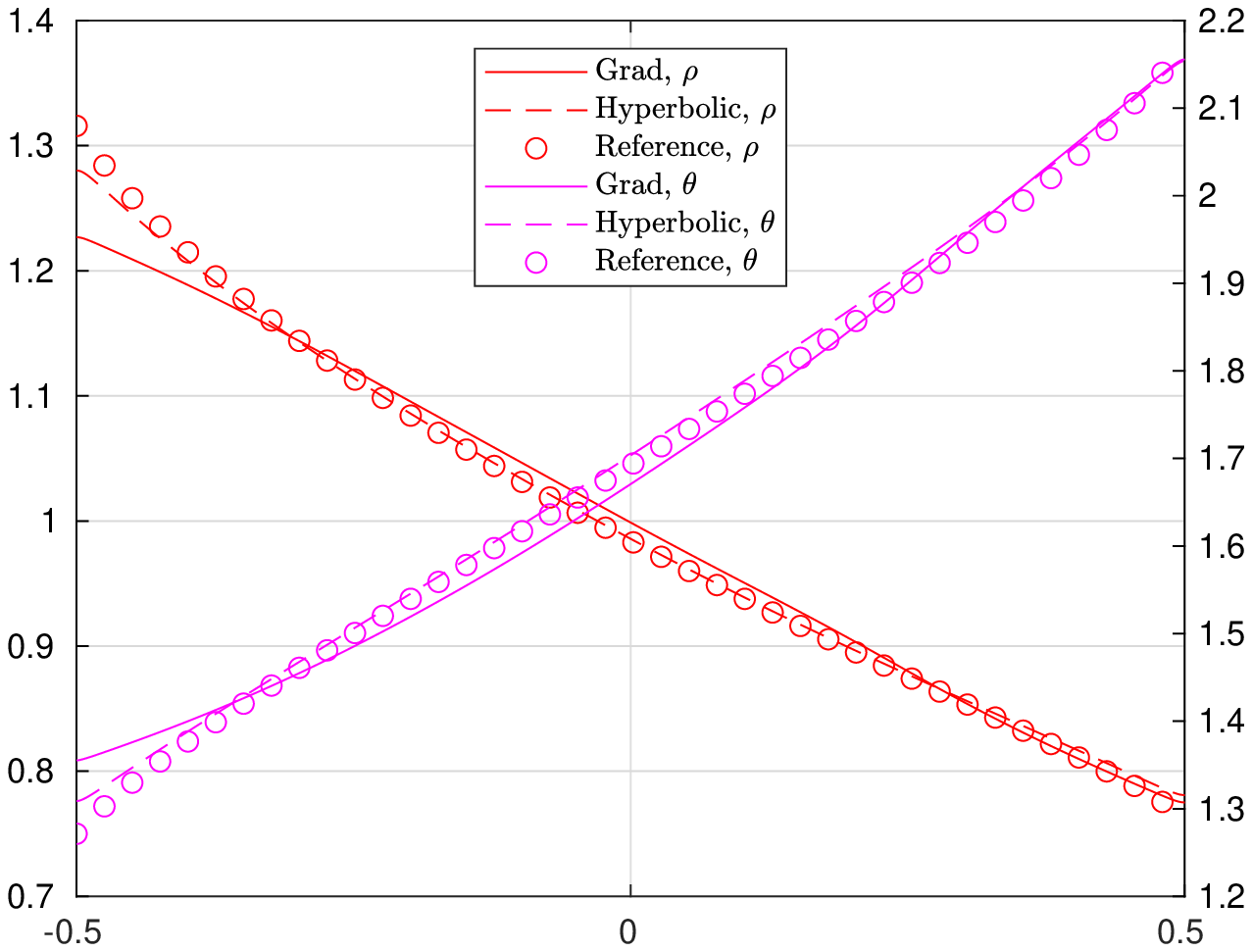}}
\hspace{10pt}
\subfloat[Phase diagram of Grad's solution\label{fig:Tr2p7_phase}]{\raisebox{-8pt}{\includegraphics[scale=.48,clip]{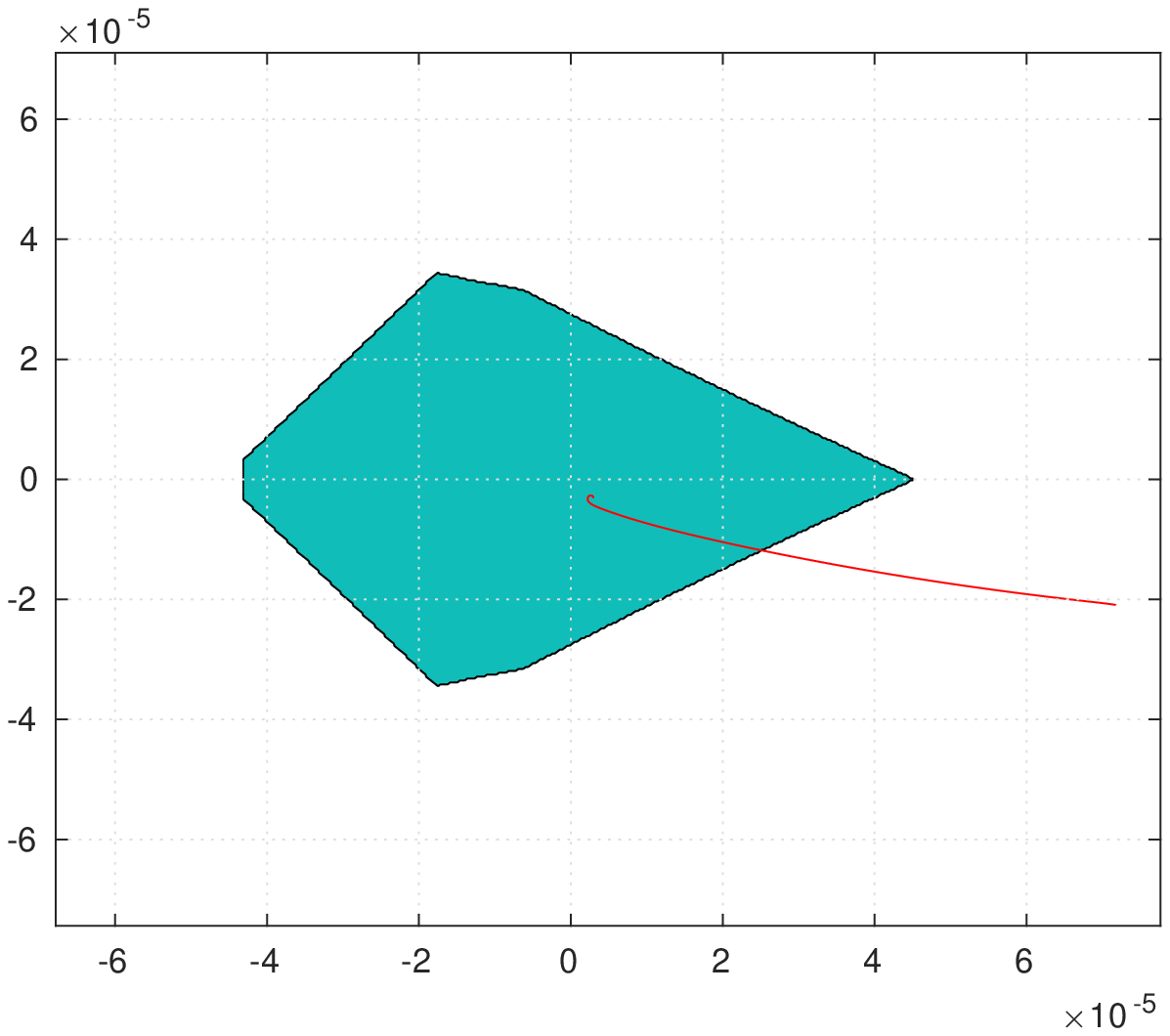}}}
\caption{Left: Steady Fourier flow for $\theta_r = 2.7$ (left vertical axis:
$\rho$, right vertical axis: $\theta$). Right: The green area is the
hyperbolicity region (horizontal axis: $\hat{f}_{M-1}$, vertical axis:
$\hat{f}_M$), and the red line is the parametric curve $(\hat{f}_{M-1},
\hat{f}_M)$ with parameter $x$.}
\label{fig:Tr2p7}
\end{figure}

For $\theta_r = 2.7$, the results are plotted in Figure \ref{fig:Tr2p7}. In
this case, if we start the simulation of Grad's equations from the initial data
\eqref{eq:init}, the computation will break down due to the loss of
hyperbolicity in the evolutional process. Therefore, we first run the
simulation for hyperbolic moment equations from the initial data
\eqref{eq:init} and evolve the solution to the steady-state. Afterward, this
steady-state solution serves as the initial data of Grad's equations. Although
the steady-state solution of Grad's equations can be found using this
technique, the approximation looks poorer than hyperbolic moment equations. The
phase diagram (Figure \ref{fig:Tr2p7_phase}) shows that the solution near the
left wall is outside the hyperbolicity region, so that the validity of boundary
conditions on the left wall becomes unclear. In contrast, the hyperbolic moment
equations still provide reliable approximation despite the high temperature
ratio.

\subsection{A summary of numerical experiments}
In all the above numerical experiments, we see that despite the loss of some
nonlinearity, the hyperbolicity fix does not appear to lose accuracy in any of
the numerical tests. In regimes with moderate non-equilibrium effects, Grad's
equations may provide solutions outside the hyperbolicity region without
numerical instability. In this situation, our experiments show that the
hyperbolicity fix is likely to improve the accuracy of the model. It has also
been demonstrated that other issues, such as subshocks and divergence, are not
related to the hyperbolicity, and these issues have to be addressed
independently.


%% file: conclusion.tex
\section{Conclusion}
The loss of hyperbolicity, as one of the major obstacles for the model reduction
in gas kinetic theory, is almost cleared through the research works in recent
years. With a handy framework introduced in Section \ref{sec:framework}, we can
safely move our focus of model reduction to other properties such as the
asymptotic limit, the stability, and the convergence issues. Our numerical
experiments show that the hyperbolic regularization does not harm the accuracy
of the model. It is our hope that such a framework can inspire more thoughts in 
the development of dimensionality reduction even beyond the kinetic theory.